\documentclass[12pt,twocolumn]{emulateapj}
\usepackage{epsf}
\usepackage{epsfig}

\def\plotancho#1{\centering \leavevmode \epsfxsize=\textwidth \epsfbox{#1}}

\begin{document}

\title{Probing the epoch of pre-reionization by cross-correlating 
cosmic microwave and infrared background anisotropies}

\author{
F. Atrio-Barandela\altaffilmark{1},
A. Kashlinsky\altaffilmark{2}}
\altaffiltext{1}{ NASA Goddard Space Flight Center, 
Greenbelt, MD 20771. On sabbatical leave from
the University of Salamanca, Spain. Email: atrio@usal.es}
\altaffiltext{2}{
Code 665, Observational Cosmology Lab, NASA Goddard Space Flight Center, 
Greenbelt, MD 20771 and
SSAI, Lanham, MD 20770; email: Alexander.Kashlinsky@nasa.gov} 

\begin{abstract}
The epoch of first star formation and the state of the intergalactic 
medium (IGM) at that time are not directly observable with current 
telescopes. The radiation from those early sources is now 
part of the Cosmic Infrared Background (CIB) and, as these sources 
ionize the gas around them, the IGM plasma would produce faint temperature 
anisotropies in the Cosmic Microwave Background (CMB) via the thermal 
Sunyaev-Zeldovich (TSZ) effect. While these TSZ anisotropies are too 
faint to be detected, we show that the cross-correlation of maps of 
source-subtracted CIB fluctuations from {\it Euclid}, with suitably 
constructed microwave maps at different frequencies can probe the 
physical state of the gas during reionization and test/constrain models 
of the early CIB sources. We identify the frequency-combined CMB-subtracted 
microwave maps from space and ground-based instruments to show that they can 
be cross-correlated with the forthcoming all-sky {\it Euclid} CIB maps to 
detect the cross-power at scales $\sim 5'-60'$ with the signal/noise of up 
to $S/N\sim 4-8$ depending on the contribution to the Thomson optical 
depth during those pre-reionization epochs ($\Delta \tau\simeq 0.05$) 
and the temperature of IGM (up to $\sim10^4$K). Such a measurement would 
offer a new window to explore emergence and physical properties of these 
first light sources.
\end{abstract}


\section{Introduction.}
Within the framework of the standard cosmological model it is theoretically 
believed that the first objects in the Universe were dominated by massive 
stars and the first black holes which formed in minihalos of $\sim10^6-10^7$M$_\odot$ 
\citep{bromm}. These objects cannot be detected individually with present-day 
instruments, so alternative methods are needed to probe the epochs of 
first star formation. 

At near-IR wavelengths (below $\sim 5\mu$m) the cosmic infrared background 
(CIB) probes emissions by stellar and black hole sources \citep{k2005} and its 
fluctuations may contain a substantial component from the first sources  
\cite{kagmm, cooray}. Significant CIB fluctuations have been identified
in {\it Spitzer} \citep{kamm1,kamm2,k12} and {\it Akari} \citep{matsumoto} data
between $\sim 2$ and $\sim 5 \mu$m after known galaxy populations have 
been subtracted to deep levels. The fluctuations are too large and have the 
wrong spatial distribution to be explained by known galaxy populations 
\citep{kamm1,kari1}. Their origin was suggested to lie at the epoch of first stars 
\citep{kamm1,kamm3} as implied by the slope of their spatial spectrum 
now measured to $\sim 1^\circ$ \citep{k12}. The high-$z$ interpretation of 
the residual fluctuations is supported by the observed lack of correlation 
between the {\it Spitzer} CIB fluctuations and optical {\it HST} image data 
out to 0.9$\mu$m where the Lyman-break at wavelength $0.1(1+z)\mu$m is 
expected \citep{kamm4}. The source-subtracted CIB fluctuations appear highly 
coherent with the soft $[0.5-2]$KeV unresolved X-ray cosmic background (CXB) 
with no detectable 
coherence appearing at harder X-ray energies \citep{c13}. The measured
correlation requires a proportion of accreting black holes among the CIB 
sources significantly higher than in the directly observed populations \citep{kari2},
further supporting the high-$z$ origin of these sources and inconsistent with the
alternative proposal for the origin of the CIB fluctuations in intrahalo light at more recent epochs 
\cite[][see also Sec. 4.3 of Helgason et al. 2014 for summary of further observational
difficulties of this scenario]{ihl}.

We will assume here that the sources producing these CIB fluctuations at $(2-5)\mu$m
lie at early epochs, $z{>\atop\sim} 8$, consistent with all the CIB-related measurements.
This will be tested definitively with the {\it Euclid} data, where this team 
has been selected by NASA/ESA to conduct an all sky near-IR CIB program 
LIBRAE (Looking at Infrared Background Anisotropies with {\it Euclid}). 
At high-$z$, these early sources would have ionized and heated up the 
surrounding gas which, in principle, would generate secondary anisotropies in
the Cosmic Microwave Background (CMB) via the thermal Sunyaev-Zeldovich (TSZ)
effect. Given that {\it Euclid} will cover $\sim 20,000$ deg$^2$ with sub-arcsec resolution at three near-IR 
channels \citep{euclid} this weak signal may be teased out of the noise, after 
suitable construction of a comparably large-area, low noise, multifrequency 
CMB maps of $\sim$arcmin resolution which are expected to be available 
in the near future. We show here how such measurements can lead to a highly 
statistically significant result for plausible modes 
of the high-$z$ evolution. At the same time, if the signal originates 
at high-$z$, there should be no correlation between the CMB and the 
diffuse emission maps obtained from the {\it Euclid} VIS (visible) channel. 
Our goal here is not to test specific models, but rather to demonstrate how the 
first ionization sources very generally can produce a measurable CIB$\times$CMB signal, 
whose detection or upper limit can then be used to probe the emergence
of the Universe out of the ``Dark Ages". The proposed technique offers to 
probe the beginning of the reionization process in a manner alternative to 
HI 21 cm tomographic studies (see Furlanetto et al. 2006 for review), both 
methods being complementary but subject to different foregrounds and systematics.

\section{CIB sources and cross-power with CMB}
\label{sec:cib-tsz}

The measured CIB fluctuation spectrum appears consistent with the high-$z$ 
$\Lambda$CDM clustering and is the same (within uncertainties) 
in different sky directions consistent with its cosmological origin \citep{k12}. 
The source-subtracted CIB fluctuations have two components: 1)
shot/white noise from the remaining (unresolved) sources dominates 
small angular scales and sets an {\it upper} limit on the shot-noise of the new 
high-$z$ populations, and 2) a fluctuation component due to clustering
of {\it new} population sources is found at scales $> 20^{\prime\prime}$. 
We use these measurements to normalize the TSZ anisotropy to the flux from, and 
the abundance of, the sources at high-$z$ using the outline similar to \cite{kamm3}.

Massive Population III stars of mass $m_*$ would radiate at the Eddington 
limit with luminosity $l_*\propto m_*$. They would have approximately constant 
surface temperature 
$T_*\sim 10^5$K and would have produced a large number of ionizing photons with
energy $\geq 13.6$eV. These lead to a constant ratio of the ionizing photons 
per H-burning baryon in these objects. The results of \cite{bromm1,schaerer} 
give a number of ionizing photons ${\cal N}_i \sim 10^{62} M_*/M_\odot$ produced 
over the lifetime of these stars ($\sim3\times10^6$yr) by a halo containing $M_*$ 
in such sources. If $\kappa$ ionizing photons are required to ionize a H atom, 
around each halo containing $M_*$ in stars there will be a bubble of 
$M_{\rm ion} \sim 10^5 \kappa^{-1}M_*$ ionized gas, heated to a temperature of
$T_e\equiv T_{e,4} 10^4$K. Hereafter we will ignore spatial variations on $\kappa$, 
and assume that most photons escape from the halos where they originate;
see discussions of \cite{santos,sf06} for stars and \cite{yue13,mesinger} 
for early black holes. If the electron temperature $T_e$ and density $n_e$ 
are constant, the Comptonization parameter averaged over the solid angle 
$\omega_B$ subtended by the bubble would be 
$Y_{C,B}=(4/3)\sigma_Tn_eR_{\rm ion}(kT_e/m_ec^2)$, 
where $R_{\rm ion}$ is the radius of the ionized cloud and $m_{e}$ 
the electron mass. Each ionized bubble would generate a CMB mean distortion 
over a pixel of solid angle 
$\omega$ given by $t_{\rm TSZ,B}=G_\nu Y_{C,B} \frac{\omega_B}{\omega}T_{\rm CMB}$
where $G_\nu$ is the frequency dependence of the effect \cite{birkinshaw}.
The net distortion will be the added contributions of all bubbles
in the CMB pixel,
$T_{\rm TSZ}= n_2\omega t_{\rm TSZ,B}$, where $n_2\omega$ is the total number
of bubbles along the line of sight on a pixel of solid angle $\omega$. 

Strong {\it lower} limits on the projected number density of sources/bubbles 
can be set by the combination of the measured {\it upper} limit on the CIB 
shot-noise power, $P_{\rm SN}$, and the amplitude of the clustering component 
of the CIB fluctuations \citep{kamm3}. Populations at high-$z$, which are 
strongly biased and span a short period of cosmic time, are expected 
to produce $\Delta \sim 10\%$ relative CIB fluctuations on arcmin 
scales requiring a net flux of $F_{\rm CIB}=\delta F_{\rm CIB}/\Delta\sim 1$ 
nW/m$^2$/sr from these populations. At the same time, since 
$P_{\rm SN} \sim F_{\rm CIB}^2/n_2$, the measured 
{\it upper} limit of $P_{\rm SN}\simeq 10^{-11}$nW$^2$/m$^4$/sr \citep{kamm2} 
implies a {\it lower} limit on the 2-D angular sky density of 
these sources of $n_2{>\atop\sim}10^{11}$sr$^{-1}P^{-1}_{{\rm SN},-11}$ 
with $P_{\rm SN}\equiv P_{{\rm SN},-11} 10^{-11}$nW$^2$/m$^4$/sr being 
the shot-noise power of the high-$z$ sources. Then, 
\begin{eqnarray}
T_{\rm TSZ} & \simeq
& \frac{4}{\pi}G_\nu T_{\rm CMB} \frac{k_BT_e}{m_ec^2}
\frac{\sigma_T}{d_A^2}\frac{M_{\rm ion}}{\mu m_H}\frac{F^2_{\rm CIB}}{P_{\rm SN}} \simeq
\nonumber \\
 & & 200G_\nu\left(\frac{0.5{\rm Gpc}}{d_A}\right)^{2}
\frac{M_*}{10^4\kappa\mu M_\odot}T_{e,4}
\frac{F_{\rm CIB}^2}{P_{{\rm SN},-11}} \; {\rm nK}
\label{eq:tsz_bubble}
\end{eqnarray}
Here, $F_{\rm CIB}$ is the net CIB flux from these sources in nW/m$^2$/sr, $\mu$
is the mean gas molecular weight and $k_B$ the Boltzmann constant. $M_*$ 
corresponds to a conservative choice for the mass of the ionizing sources 
in each early halo. In standard cosmology $d_A$=0.5--0.9 Gpc at $z$=20--10. 
For the parameters of eq. \ref{eq:tsz_bubble}, the effective Thomson optical 
depth due to the reionized medium is 
$\tau_{\rm eff}\equiv200 {\rm nK}/[T_{\rm CMB}(k_BT_e/m_ec^2)]=0.044$, well below
the measured value of $\tau=0.097\pm 0.038$ \citep{planck16}. 

Due to the variation on the number density of bubbles with a relative number
fluctuation of $\Delta\simeq 0.1$ \citep{kamm3}, the CMB distortion $T_{\rm TSZ}$ would 
generate CMB temperature fluctuations. If we further assume that the ionized 
bubbles are distributed as the halos where the ionized photons originate, the 
TSZ temperature anisotropies would have an amplitude $\sim T_{\rm TSZ}\Delta$
that could potentially be detected by cross-correlating the produced CMB 
anisotropies with CIB fluctuations. If $P_{\rm CIB}$, $P_{TSZ}$
and $P_{\rm CIB \times TSZ}$ are the power spectrum of the CIB flux, TSZ
anisotropies and their cross-power, respectively, the coherence between
CIB sources and bubbles is $C=P_{\rm CIB \times TSZ}^2/(P_{\rm CIB}P_{\rm TSZ})$. 
For bubbles coherent with CIB sources ($C\simeq 1$),
the cross-power between CIB and TSZ is $P_{\rm CIB \times TSZ} 
\simeq \sqrt{P_{\rm CIB}}\sqrt{P_{\rm TSZ}}$. To compute this
cross-correlation, the sub-arcsec {\it Euclid} CIB 
and arcmin-resolution CMB maps will be brought to a common 
resolution. When measuring the cross-power from 
IR and microwave maps ($\mu$) of $N_{\rm pix}$ CMB pixels, the error is 
$\sigma_{P_{\rm CIB \times TSZ}}\simeq \sqrt{P_{IR}}\sqrt{P_\mu}/\sqrt{N_{\rm pix}}$
\citep{k12,c13}. At the scales of interest here (${>\atop\sim} 1^\prime$) the 
{\it Euclid} CIB maps will have negligible noise, so $P_{IR}=P_{\rm CIB}$. Using the {\it Euclid} Wide Survey
of ${\cal A}\simeq 20,000$deg$^2$ the CIB power on arcmin scales will
be measurable by LIBRAE down to sub-percent statistical accuracy. 
If primary CMB is removed, the foreground-reduced
microwave maps would be dominated by instrument noise $\sigma_n$, foreground
residuals $\sigma_{f,res}$ and, more importantly, the TSZ of the unresolved 
cluster population $\sigma_{unr,cl}$. With $N_\nu$ frequency channels the 
variance of the microwave map would be 
$\sigma_\mu^2=\sigma_n^2/N_\nu+\sigma_{f,res}^2+\sigma_{unr,cl}^2$.
The signal-to-noise would be
S/N$\simeq T_{\rm TSZ}\Delta\sqrt{N_{\rm pix}}/\sigma_\mu$, which can reach 
S/N$\gg 1$ for some experimental configurations discussed below. Specifically
\begin{equation}
{\rm S/N} = 7 \;\;\frac{T_{\rm TSZ}}{200nK}\;\frac{\Delta}{0.1}\;
\frac{5\mu\rm K}{\sigma_\mu}\left(\frac{N_{\rm pix}}{3\times10^6}
\right)^{\frac{1}{2}}\;,
\label{eq:s2n}
\end{equation}
where $N_{\rm pix}=3\times10^6$ corresponds to 20,000 deg$^2$ binned into 
independent squares of $5^\prime$ on the side, the expected sky coverage
of the {\it Euclid} satellite downgraded to the native resolution of {\it Planck}.

\section{CMB data}
\label{sec:cmb}

Eq.~\ref{eq:s2n} shows that CMB maps with low $\sigma_\mu$
covering a sufficiently large sky area are needed for a statistically significant 
measurement of the CIB$\times$TSZ cross correlation. Such maps already exist
and more will be available in time for the upcoming {\it Euclid} CIB 
measurement with LIBRAE. Resolved sources in {\it Spitzer} CIB analysis remove
$\sim 25\%$ of the sky \citep{kamm1,k12}; this fraction is expected to be lower for the much
better resolution of {\it Euclid}, which in any event will then be pixelated into the CMB
resolution of $>\atop\sim$arcmin.

To  transform these CMB 
data into maps suitable for probing the contribution of the TSZ from
the first stars era and gain insight into the epochs prior to completion 
of the reionization, one needs to remove the 
primary CMB component of $\sigma_{\rm CMB} \sim 80 \mu$K, which is 
feasible with the multi-frequency microwave maps obtained, and obtainable, 
with numerous instruments by, for example, subtracting a 217GHz map, the
null of the TSZ effect, from maps at other
frequencies.  In this process other black-body components, 
such as from the integrated Sachs-Wolfe and kinetic SZ effects, will 
also be removed. After the suitable CMB subtraction has been achieved 
the microwave maps will be cross-correlated, via the cross-power \citep[see][]{k12,c13}, 
with the {\it Euclid} CIB maps to be produced over patches of $\sim 1^\circ$ 
on the side and covering 20,000 deg$^2$ in the {\it Euclid} Wide Survey. 
We now specify several instrumental configurations for such a measurement.

$\bullet$ {\it Planck}. 
For {\it Planck} channels at frequencies $\nu=(100, 143, 217,353)$GHz, the beams 
have an effective FWHM=$(9.65,7.25,4.99,4.86)^\prime$ \citep{planck6}. At these 
frequencies $G_\nu=(-1.51,-1.04,0.,2.23)$ and the noise after the planned two years 
of integration is $\sigma_n=[1.3, 0.5, 0.7, 2.5]\mu$K on pixels of $1^\circ$ side.
Other contributions to the overall $\sigma_\mu$ come from sources below 
the threshold detection level of the instrument. The amplitude of the 
Poisson and clustered foreground power spectra 
$D_\ell=\ell(\ell+1)C_\ell/2\pi$ at $\ell=3000$
are $A^{PS}=(220\pm 53,75\pm 8,60\pm 10)\mu$K$^2$ and
$A^{CL}=(\sim 0, 32\pm 8, 50\pm 5)\mu$K$^2$ for 100, 143 and 217GHz, after 1 
yr of integration \citep{planck15}. The Planck Collaboration does not provide 
values of these contributions at 353GHz, so we will assume them to be negligible.
These components will not cancel when subtracting the 217GHz map from 
those of other frequencies. The TSZ from unresolved
clusters has an amplitude $A^{TSZ}=4G_\nu^2\mu$K$^2$, that will 
also be present \citep{planck16}. In summary, the noise in the {\it Planck} CMB 
difference maps $\nu-217$GHz is $\sigma_{\nu-217GHz}=(9.6,10.3,32.0)\mu$K for
$\nu=(100,143,353)$GHz on pixels corresponding to the FWHM at each frequency. Combining the
three channels, taking into account the frequency dependence of
the TSZ effect, the error on the Comptonization parameter
would be $\sigma_{\mu}^{Planck}=7.8\mu$K on pixels of 5$'$ after 2 yrs. 

$\bullet$ {\it The Atacama Cosmology
Telescope} (ACT) has already mapped CMB over $600$deg$^2$ at 148, 218
and 277GHz  with resolution $(1.4^\prime,1.0^\prime,0.9^\prime)$ \citep{act}. 
The currently observed area is small for the purposes of the present discussion
and adds little to the S/N of the measurement, but an expanded area 
could eventually provide useful data.

$\bullet$ {\it South Pole Telescope} (SPT) has already mapped 2,540 deq$^2$
at (95, 150, 220)GHz, with resolution (1.7, 1.2, 1.0)$^\prime$. The
noise levels are 18$\mu$K at 150GHz and $\sim\sqrt{2}$ larger for the 
other two channels \citep{spt}. The CMB-subtracted maps have a residual
noise of $\sigma_{\nu-220GHz}\simeq (36,32)\mu$K with a similar number
of pixels as Planck ($N_{pix}>3\times 10^6)$ for the lowest resolution 
channel. Combining the two frequencies, the noise on the
Comptonization parameter scaled to 5$'$ pixels is $\sigma_\mu^{SPT}=4.74\mu$K,
smaller than that of Planck but over 1/8 the area covered by {\it Euclid}. 
It would provide competitive results with
those of {\it Euclid} and {\it Planck}. In addition, due to its location, the SPT data
should be useful for a similar analysis with WFIRST \citep{wfirst}.

$\bullet$ {\it Future experiments} like Advanced ACTPol \citep{actpol}
and CMB-Stage 4 \citep{fourthg} would aim to map 1/2-2/3 of the sky
with sensitivity of $\sim 10\mu$K/arcmin and $\sim 1\mu$K/arcmin
at a wide range of frequencies, 30-300GHz and 40-240GHz, respectively.
In these future instruments, the limiting factor will not be instrumental
noise, but confusion from foreground sources and TSZ contributions from
unresolved clusters. These foregrounds 
would contribute $\sigma_{for}\sim 3-10\mu$K in the frequency range 100-217GHz.
These low noise levels, in combination with {\it Euclid} and, potentially also 
WFIRST-based, CIB maps, offer the exciting possibility of probing how CIB
sources affected the physical properties of the IGM prior to reionization.
 
\section{The CIB-CMB cross power spectrum}

We can refine the previous estimate (eq.~\ref{eq:s2n}) by a specific model
computation of the cross-power of TSZ anisotropies and CIB fluctuations.
We compute the TSZ-CIB cross-power on angular scale $\theta=2\pi/q$ via 
the relativistic Limber equation
\begin{equation}
P_{CIB\times TSZ}(q)=\int dr(z)\left(\frac{dF_{\rm CIB}}{dr}\right)
\left(G_\nu\frac{dY_C}{dr}\right) \frac{P_{3}(q/r,z)}{r^2} \,,
\label{eq:cib_tsz}
\end{equation}
and normalize it to the {\it measured} CIB auto-power as discussed 
in \citep{jwst}. In the above, 
$q=2\pi/\theta$, $P_{3}$ is the 3-D biased power spectrum of the sources, and 
$r(z)=d_A(1+z)=d_L/(1+z)$ is the comoving distance. 

As in \cite{kagmm}, we assume that the sources radiate at the Eddington 
limit, form proportionally to their collapse rate and, in addition, 
are designed to reproduce the measured {\it integrated} CIB fluctuations 
over (2-5)$\mu$m, verifying that their relative fluctuation $\Delta\simeq 0.1$ 
at $2\pi/q=5'$ \citep{jwst}. As the ionized bubbles appear around 
the CIB sources, they will start generating the optical depth due to 
Thomson scattering, $d\tau/dr = \rho_{\rm baryon}/(\mu m_p)
\sigma_T x_e(1+z)^{-1}$ where $x_e$ is the ionization fraction, as well 
as contribute to the Comptonization, $dY_c/dr=(k_BT_e/m_ec^2)d\tau/dr$. 
Because the ionized bubbles surround the first sources, we assume the same
biasing for both TSZ and CIB and use the methodology of \cite{kagmm,cooray} 
to relate it to the underlying standard $\Lambda$CDM power spectrum.

In principle, the ionization fraction $x_e(z)$ can 
be related to fraction of baryons in stars as: $x_e=n_If_*$, where $n_I$
is the number of ionized atoms per baryon in CIB sources and $f_*(z)$ 
is the fraction locked in CIB sources up to $z$. To simplify our treatment, 
we parametrize $f_*(z)\propto {\rm erfc}((z-z_0)/\Delta z)$
where $z_0$ is the redshift at which half of $f_*$ have been locked up in
CIB sources, $\Delta z$ determines how fast baryons collapse to form CIB sources.
We normalize below the overall optical depth produced by these sources to 
$\Delta \tau=0.05$ and take the temperature of the gas in the 
bubbles to be $T_{e}=10^4$K, where atomic cooling becomes inefficient 
in the absence of H$_2$. In the presence of heating by accreting  
black holes or X-ray binaries, $T_e$ can rise somewhat above $10^4$K \citep{mirabel11,jeon}. 

\begin{figure*}
\centering
\plotancho{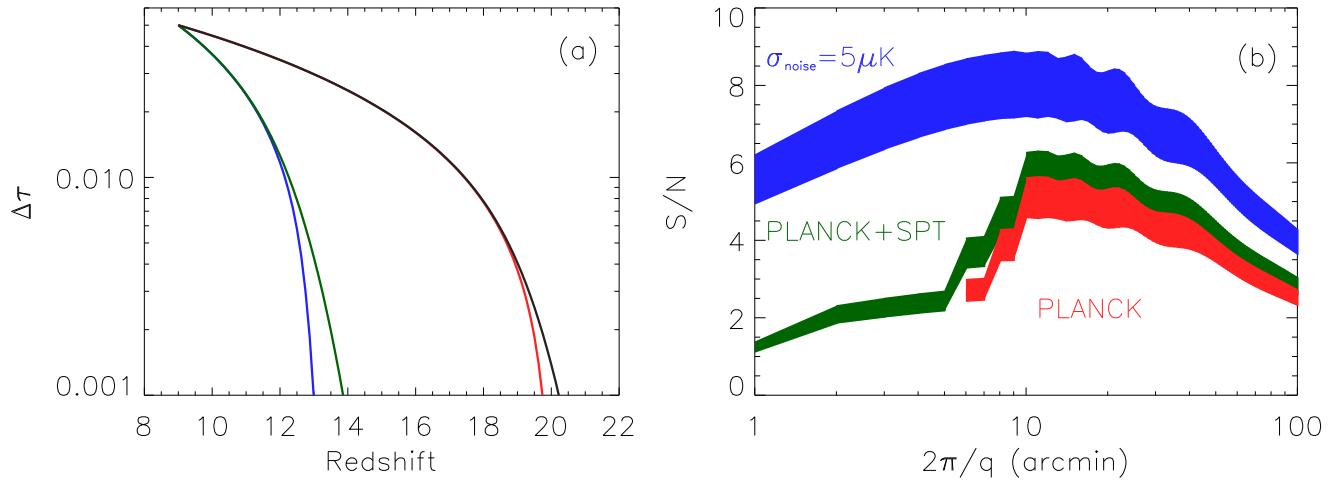}
\vspace*{-5cm}
\caption[fig:fig1]{{\it (a)}  The build-up of $\Delta \tau$ for the 
models considered here. Blue/green lines correspond to $\Delta z=0.3/1.3$ 
and $z_0=13$, and red/black to $\Delta z=0.3/1.3$ and $z_0=20$.
{\it (b)} Filled regions show the range of the S/N of the CIB-TSZ cross power 
over the {\it Euclid} Wide Survey region covered by the model parameters for 
different experimental configurations. {\it Planck} parameters are projected 
to 2 yrs of integration and include terms discussed in Sec. \ref{sec:cmb}.
At 5$'$ only 353--217 GHz difference maps would be useful, at 7$'$ we 
also add 143--217 GHz, and at $>9'$ we can add the data from 100--217 GHz.
SPT has lower S/N, but can probe angular scales as low as $\sim 1'$. In 
its current configuration the ACT does not add appreciably to the 
measurement, but that can be improved with Advanced ACT and CMB Stage 4 
experiments as shown with the blue band.
}
\label{fig:fig1}
\end{figure*}

We assumed in computations that $z_0=[13,20]$ and $\Delta z=[0.3,1.3]$ 
fixing the final ionization fraction to reproduce $\Delta \tau =0.05$; 
this corresponds to $0.3<x_e<1$ at the end of that period.
We verify that the coherence of these models varies between 
$0.8<\sqrt{{\cal C}_{\rm TSZ\times CIB}}<0.95$ to compare with the discussion 
in Sec. \ref{sec:cib-tsz}. In Fig~\ref{fig:fig1}a we show how the optical 
depth reaches $\Delta\tau=0.05$ for these models. Fig~\ref{fig:fig1}b shows
S/N with which the experiments discussed in Sec. \ref{sec:cmb} can probe the 
CIB$\times$TSZ cross-correlation at different angular scales. We always 
assume that the experiment has at least two frequencies with overlapping 
resolution so the primary CMB and the kinetic SZ component can be removed from the 
maps. The band width encloses the minimum/maximum values of all our models.  
Notice that the final S/N is weakly dependent on when reionization occurs
(parametrized by $z_0$) and the width of this period ($\Delta z$). This is a consequence
of normalizing all models to the measured CIB flux and to the same
electron optical depth. The blue band corresponds to an experiment with the
instrumental noise and foreground residuals of $\sigma_{noise}=5\mu$K on
pixels of 5$'$ side covering the area of the {\it Euclid} wide field survey.
The red band corresponds to the S/N using {\it Euclid} and the HFI 
{\it Planck} data at the end of the nominal 2yr mission. The S/N
is $\sim 2$ when using the 353GHz and 217 GHz channels and increases to 
$\sim 5$ when the 143 and 100 GHz channels are added. Adding SPT data 
\citep{spt} increases the S/N to $\sim 6$ at 13$'$.

The results of Fig~\ref{fig:fig1} scale as $(S/N)\propto \Delta\tau\cdot T_e$ leading to higher
significance for $T_{e,4}>1$ as reached in the modeling of e.g. \cite{jeon}.
The S/N can be increased by computing the cross-power over wider angular bins
allowing to probe much lower parameters as done in \cite{c13}.  In addition, in 
experiments such as {\it Planck}, observing at frequencies above and below 
the TSZ null frequency, the CIB-TSZ correlation will change sign, 
offering a potential test to eliminate spurious contributions.  

\section{\label{sec:conclusions}Discussion}

We have demonstrated that the cross-correlation of the future CIB {\it Euclid}
maps with the CMB data expected by that time can probe or constrain the 
pre-reionization epochs as the Universe emerges out of the ``Dark Ages". 
If 1) these sources contributed about half of the measured optical depth 
to the last scattering surface and 2) the ionized blobs contained gas at 
$T_e=10^4$K, the measurement would be at S/N$>6$ using the CMB map 
differencing method discussed here. The cross-power measurement  can be 
further improved by computing the latter in wider angular bins. The 
overall S/N demonstrated here is large enough for the detection to 
remain significant even if parts of the {\it Euclid} CIB maps 
are polluted by Galactic cirrus, which would not correlate with the 
TSZ component. 

At fixed angular resolution the significance scales with the area of CIB-CMB overlap, $A$, as 
S/N $\propto \sqrt{A}$. Additionally, WFIRST, the Dark Energy mission currently 
considered by NASA, could likewise provide significant results with 
its coverage of four near-IR channels currently planned to cover 
$\simeq 2,000$deg$^2$ \citep{wfirst} combined with the proposed CMB-Stage 4 
arcmin-resolution instrument \citep{fourthg}. The planned absence of the visible 
channel on-board WFIRST in testing the high-$z$ origin of the 
cross-power can be compensated with the {\it Euclid} VIS data. 

{\bf Acknowledgements:} NASA's support to the LIBRAE project (PI, A. Kashlinsky) 
NNN12AA01C is gratefully acknowledged.
FAB acknowledges financial support from the Spanish
Ministerio de Educaci\'on y Ciencia (project FIS2012-30926). We thank 
Eric Switzer for useful information on current and future CMB experiments and 
Rick Arendt and Ed Wollack for comments on the draft manuscript.

\end{document}